\documentclass[sigconf, nonacm]{acmart}

\usepackage{gensymb}

\usepackage{enumitem}

\newcommand{\sysname}{GlanceAR}

\AtBeginDocument{%
  }

\setcopyright{acmlicensed}
\copyrightyear{2026}
\acmYear{2026}
\acmDOI{XXXXXXX.XXXXXXX}

\acmConference[UIST '26]{ACM Symposium on User Interface Software and Technology}{November 02--05, 2026}{Detroit, MI}

\acmISBN{978-1-4503-XXXX-X/2026/11}

\begin{document}

\title{\sysname{}: Repurposing Phone Sensors via Low-Cost Reflectors for Always-On Obstacle Awareness}

\author{Boyang Zhou}
\orcid{1234-5678-9012}
\affiliation{%
  \institution{University of Washington}
  \city{Seattle}
  \state{WA}
  \country{USA}
}
\email{zby2003@cs.washington.edu}

\author{Xizhe Hao}
\affiliation{%
  \institution{University of Washington}
  \city{Seattle}
  \state{WA}
  \country{USA}
}
\email{xhao6@uw.edu}

\author{Minbeom Kim}
\affiliation{%
  \institution{KAIST}
  \city{Daejeon}
  \country{Korea}
}
\email{kmb3322@kaist.ac.kr}

\author{Jaewook Lee}
\affiliation{%
  \institution{University of Washington}
  \city{Seattle}
  \state{WA}
  \country{USA}
}
\email{jaewook4@cs.washington.edu}

\author{Jon E. Froehlich}
\affiliation{%
  \institution{University of Washington}
  \city{Seattle}
  \state{WA}
  \country{USA}
}
\email{jonf@cs.washington.edu}

\renewcommand{\shortauthors}{Zhou et al.}

\begin{abstract}
Smartphone use while walking reduces environmental awareness and can lead to accidents and injury, yet existing obstacle-awareness systems remain unadopted because they require users to hold the phone upright and aimed forward, conflicting with how people actually hold their devices. We introduce \textit{\sysname{}}, an always-on mobile obstacle-awareness system designed for natural phone-holding postures. \sysname{} utilizes a low-cost passive mirror attachment to redirect the phone's rear LiDAR and RGB camera forward without additional electronic components. In a within-subjects study with $N{=}11$ participants navigating indoor and outdoor obstacle courses while engaged in a primary on-phone task, we evaluated three sensing configurations: \sysname{}, a LiDAR-only variant for privacy, and an attachment-free ultra-wide camera baseline. Our findings reveal trade-offs among sensing fidelity, obstacle cue design, and social acceptability, and surface insights around AI complacency and the tension between on-screen task focus and physical safety.
\end{abstract}



\keywords{pedestrian safety, obstacle detection, computer vision, LiDAR}
\begin{teaserfigure}
\centering
  \includegraphics[width=\textwidth]{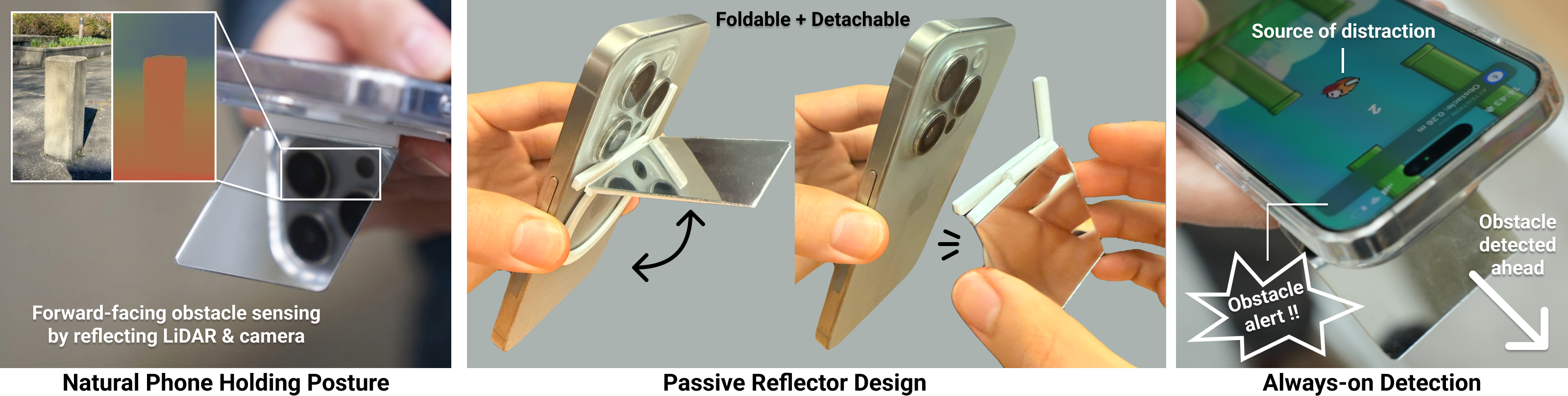}
  \caption{\sysname{} is a mobile obstacle-awareness system that uses a passive mirror attachment to redirect a phone's rear sensors forward, delivering hazard alerts over applications while allowing users to maintain a natural phone-holding posture.}
  \Description{}
  \label{fig:teaser}
\end{teaserfigure}


\maketitle

\section{Introduction}
Smartphone use while walking is a well-documented safety hazard~\cite{nasar2008mobile, hyman2010did}---smartphone-related pedestrian injuries have more than doubled in a decade~\cite{scopatz2016effect}. Existing digital interventions remain inadequate~\cite{arafat2023effectiveness}, with no solution having achieved mainstream adoption. Emerging research prototypes such as \textit{WalkSafe}~\cite{wang2012walksafe} and \textit{BumpAlert}~\cite{tungUsePhoneSensors2018} demonstrate that phone sensors can support obstacle detection using rear-camera vision and acoustic ranging. However, these systems all assume the phone is oriented with its rear sensors facing the walking path, conflicting with the natural downward-tilted posture of everyday phone use~\cite{jackermeierUserBehaviourWild2019}.



We introduce \textit{\sysname{}}, an always-on, on-device obstacle awareness system designed for natural phone-holding postures. The core challenge is that in this posture, the head tilts 30--40$^{\circ}$ downward~\cite{han2019head} and the phone even further~\cite{jackermeierUserBehaviourWild2019}, pointing the rear camera and LiDAR at the ground rather than the path ahead. \sysname{} addresses this with a passive mirror on a 3D-printed mount (\$0.30) that redirects the phone's rear LiDAR and camera into a forward-facing orientation while the user maintains a natural grip.
When the reflected LiDAR distance drops below a threshold, an on-device VLM (\textit{FastVLM} 0.5B~\cite{vasu2025fastvlmefficientvisionencoding}) identifies the obstacle semantically (\textit{e.g.,} pedestrian, bollard) and delivers multi-modal, glanceable cues layered over the user's on-screen task. To our knowledge, \sysname{} is the first system to enable forward-looking obstacle awareness during natural phone use without additional electronic components, and the first to combine passive optical redirection of commodity LiDAR with on-device VLM-powered hazard detection. The current mirror is admittedly bulky, but we refined it into a foldable design based on participant feedback, and accessories like collapsible phone stands and periscope lenses suggest even slimmer forms are feasible.



We evaluate \sysname{} in a within-subjects study ($N{=}11$) across three sensing configurations: (1) the \textit{primary \sysname{} mirror attachment} covering both LiDAR and RGB camera; (2) a smaller \textit{LiDAR-only mirror ablation} that avoids capturing forward-facing imagery for privacy; and (3) an \textit{attachment-free technical baseline} that exploits the 120$^{\circ}$ field of view of the ultra-wide camera to capture the walking path $\sim$1.5\,m ahead even in a downward-tilted posture. Each configuration offers different tradeoffs across sensing fidelity, hardware modification, and privacy.
Our results suggest physical sensing configurations had little effect on look-up rates; rather, participant behavior and perceived safety were driven almost entirely by the simplicity and low information density of the obstacle cues.
Building on these findings, we discuss how \sysname{} surfaces tension between offloading environmental awareness to an automated system and encouraging unsafe phone use, alongside bystander privacy concerns introduced by redirected mobile sensing.

In summary, our contributions include: (1) \textit{\sysname{}}, an always-on obstacle awareness system with a passive mirror attachment to redirect commodity phone sensors for forward hazard detection during natural, downward-facing phone use; (2) a design space of visual, auditory, and haptic obstacle cues informed by user feedback; and (3) an empirical evaluation ($N=11$) of three sensing setups (full mirror, LiDAR-only, and attachment-free baseline), characterizing detection performance and surfacing insights on AI complacency, task focus versus safety, and passive sensing privacy.

\section{Related Work}

Prior work in smartphone-based pedestrian hazard detection spans computer vision, mobile computing, and accessibility. We build on this foundation with \sysname{}, extending these sensing capabilities to everyday, natural phone-holding postures.

\subsection{Hazard Detection for Pedestrians}

Several research prototypes have demonstrated that commodity phone sensors can support real-time hazard detection. \textit{WalkSafe}~\cite{wang2012walksafe} uses the rear camera to detect approaching vehicles and alert distracted users at intersections, while \textit{BumpAlert}~\cite{tungUsePhoneSensors2018} combines acoustic ranging with camera data to warn of nearby obstacles. \textit{CrowdWatch}~\cite{wangCrowdWatchDynamicSidewalk2017} takes a crowd-sourced approach, aggregating mobile sensor data to map sidewalk hazards. Beyond phones, vehicle-mounted sensor suites~\cite{fardiObstacleDetectionPedestrian2006, sewalkarVehicletoPedestrianCommunicationVulnerable2019} and dedicated phone attachments~\cite{arafatEffectivenessInterventionsMobile2023a, yeoOmniSenseExploringNovel2023, hincapie2013crashalert, ultrasound2015} have also been explored, though these require electronics beyond what pedestrians already carry.

A shared limitation across the phone-only approaches is that the device must be oriented with its rear sensors facing the direction of hazards~\cite{tungUsePhoneSensors2018, seeSmartphoneBasedMobilityAssistant2022, tapuSmartphoneBasedObstacleDetection2013, wang2012walksafe}. However, field studies of naturalistic phone use show that users tilt their heads 30–40$^{\circ}$ downward while texting or browsing~\cite{han2019head}, and the phone itself is angled further still~\cite{jackermeierUserBehaviourWild2019}, leaving rear sensors aimed well below the horizon. \sysname{} is designed around this empirical reality, enabling forward hazard detection from the postures people actually adopt rather than the upright orientation prior systems require.

\subsection{Obstacle Awareness for BLV}

Obstacle awareness has been studied extensively for blind and low-vision (BLV) users~\cite{dakopoulosWearableObstacleAvoidance2010}, ranging from dedicated wearables~\cite{rodriguezAssistingVisuallyImpaired2012, diaztoroVisionBasedSystemAssisting2021, jainReviewLiDARBasedNavigation2023, josephStateoftheArtReviewWearable2023, xuWearableObstacleAvoidance2023} to phone-based solutions~\cite{tapuSmartphoneBasedObstacleDetection2013, seeSmartphoneBasedMobilityAssistant2022, pengSmartphoneBasedObstacleSensor2010}. More recently, iOS 26's Magnifier app brings VLM-powered detection of people, doors, scenes, and furniture to iPhones~\cite{MagnifyDescribeThings}. Across these approaches, however, environmental monitoring is treated as BLV users' primary task, and devices are oriented to maximize sensing coverage. \sysname{} takes a different approach, delivering obstacle cues alongside the on-screen task to fit how people actually use their phones.

\subsection{Mirror-Based Sensor Redirection}

Passive mirrors provide a lightweight way to reshape a sensor’s field of view without modifying hardware~\cite{getschmannMirrorForgeRapidPrototyping2022}. Prior HCI systems have applied mirrors to augment a camera's field of view, enabling novel sensing techniques such as revealing otherwise-occluded surfaces for one-handed gestures~\cite{wongBackMirrorBackofdeviceOnehanded2016} or expanding the effective capture volume of on-device cameras for 3D capture and hand/body input~\cite{yuAccurate3DShape2019, yuHandSeeEnablingFull2019, ahujaMeCapWholeBodyDigitization2019, aokiFaceRecGlassesWearableSystem2021, surround-see2013}. Beyond standard RGB cameras, mirror redirection has also been applied to reshape LiDAR fields of view~\cite{kibiiDesignCalibrationPlane2022, pelkaCalibrationPlanarReflectors2021}. \sysname{} builds on these insights and applies mirror-based redirection to phone-based obstacle detection. By attaching a passive mirror, our system redirects both the RGB camera and LiDAR sensor into a forward-looking view while the phone remains tilted downwards.

\section{The \sysname{} System}
As a safety-critical system, \sysname{} operates entirely on-device on a LiDAR- and camera-equipped iPhone. Because the phone's rear sensors point at the ground in a natural holding posture (45°--90° from vertical~\cite{han2019head, jackermeierUserBehaviourWild2019}), the system uses a passive mirror to redirect them forward, continuously monitoring for obstacles and delivering real-time alerts over the user's primary on-screen task.



\subsection{Passive Reflector Design}

\begin{figure}[h]
    \centering
    \includegraphics[width=\columnwidth]{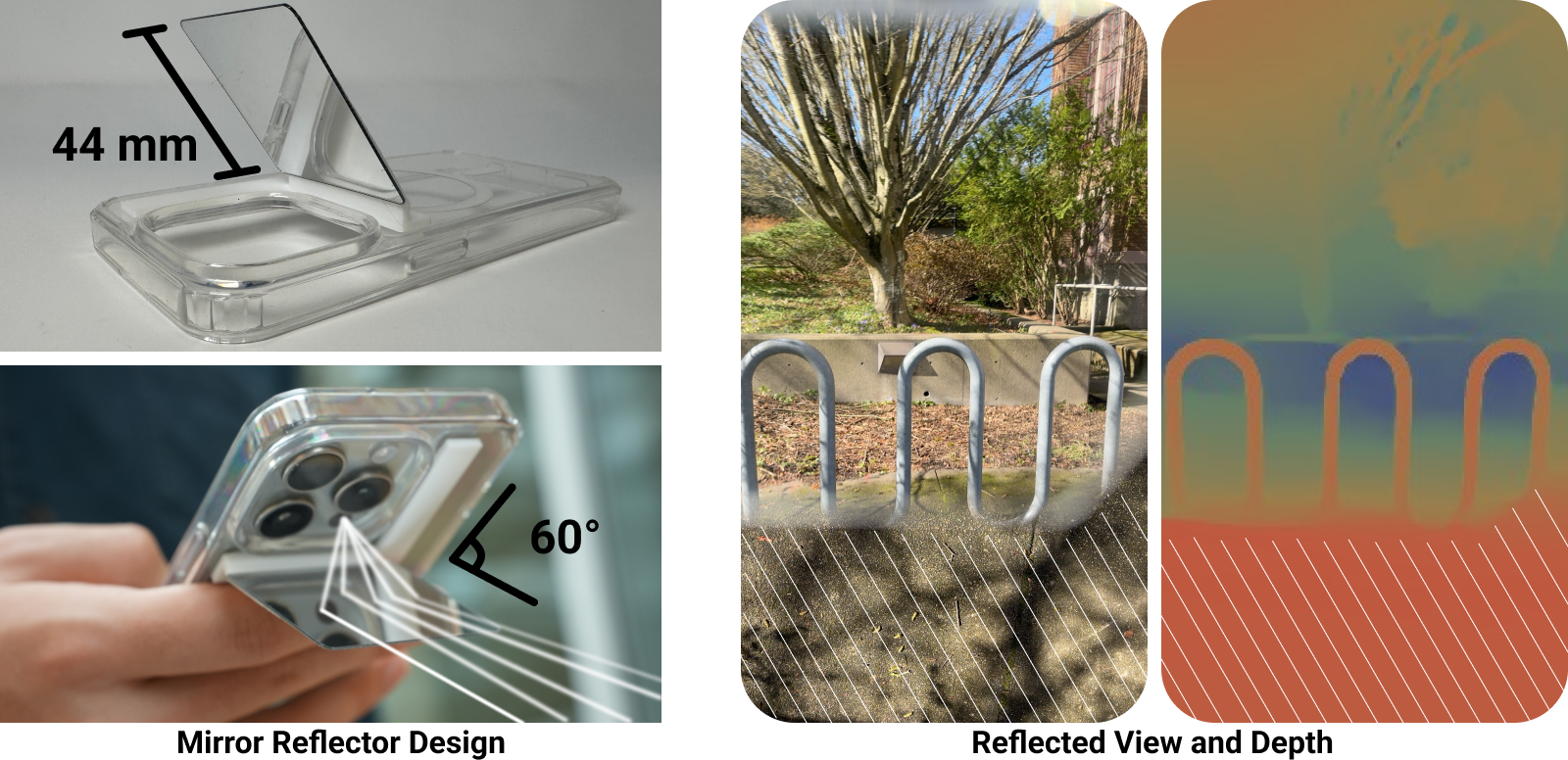}
    \caption{
    Passive reflector attachments enabling forward-looking sensing with iPhone LiDAR \& camera.
    \textit{Left:} \sysname{} Reflector attachment. 
    \textit{Right:} Reflected forward-facing RGB feed and reflected depth map after RGB–LiDAR sensor fusion. The unreflected dashed region is discarded during processing.
    }
    \label{fig:reflector}
    \vspace{-0.2cm}
\end{figure}

Modern iPhones are equipped with direct Time-of-Flight (dToF) LiDAR sensors and high-quality RGB cameras.
As noted in prior sensor research, dToF sensors adhere to basic optical reflection principles \cite{pelkaCalibrationPlanarReflectors2021}.
Building on this intuition, we developed a passive mirror reflector attachment mounted over the iPhone's camera–LiDAR module that redirects the native downward sensing direction to a forward-facing one, preserving sensing capability while the phone remains in a natural posture (\autoref{fig:reflector}, left).

The core \sysname{} reflector consists of a planar mirror and a 3D-printed support stand tilted at $60^{\circ}$.
The mirror measures 44 mm × 44 mm to cover both the LiDAR sensor and the majority of the main (1×) RGB camera (located on the bottom-left of the camera module for iPhone 15 Pro).
To ensure consistent optical alignment, the mount includes a right-angle indentation that seats against the phone’s camera bump, mechanically constraining placement so that the mirror maintains the intended reflection angle.
For everyday stability, the \sysname{} reflector is designed to be semi-permanently adhered to a dedicated phone case.
The contact base between the reflector and the phone is designed to be easily scalable to different camera-module layouts, allowing the same reflector geometry to be adapted to other phone models by adjusting the footprint dimensions.
The combined material cost of the reflector attachment is approximately \$0.30.

Finally, because the mirror does not entirely encompass the camera's vertical field of view, the raw captured frame includes an unreflected bottom region. To account for this, \sysname{} crops and discards this downward-facing area during processing, as indicated by the dashed region on the right side of~\autoref{fig:reflector}.

\subsection{Obstacle Perception Pipeline}

To process the redirected forward-facing feed, \sysname{} leverages Apple’s sensor fusion pipeline and jointly processes RGB and depth captures at 30 FPS. This configuration produces denser and more stable depth estimates than reflected LiDAR alone.

To identify the presence of obstacles, we apply simple thresholding to the metric depth measurements obtained from the sensor fusion. We isolate the reading from the phone to the closest detected point, recording the shortest distance to any obstacle ahead. 

When the distance to the closest obstacle falls below a predefined threshold, \sysname{} invokes an on-device VLM to extract open-vocabulary semantic information about the forward scene, identifying obstacle types such as trees, pedestrians, and bollards.
Specifically, we prompt an off-the-shelf FastVLM 0.5B~\cite{vasu2025fastvlmefficientvisionencoding} via Core ML.
While lightweight VLMs currently exhibit variable accuracy, this configuration enables low-latency scene understanding, averaging 767 ms to the first token on an iPhone 15 Pro.

\subsection{Obstacle Cues}
\label{section:all_cues}
\begin{figure*}[h]
    \centering
    \includegraphics[width=\textwidth]{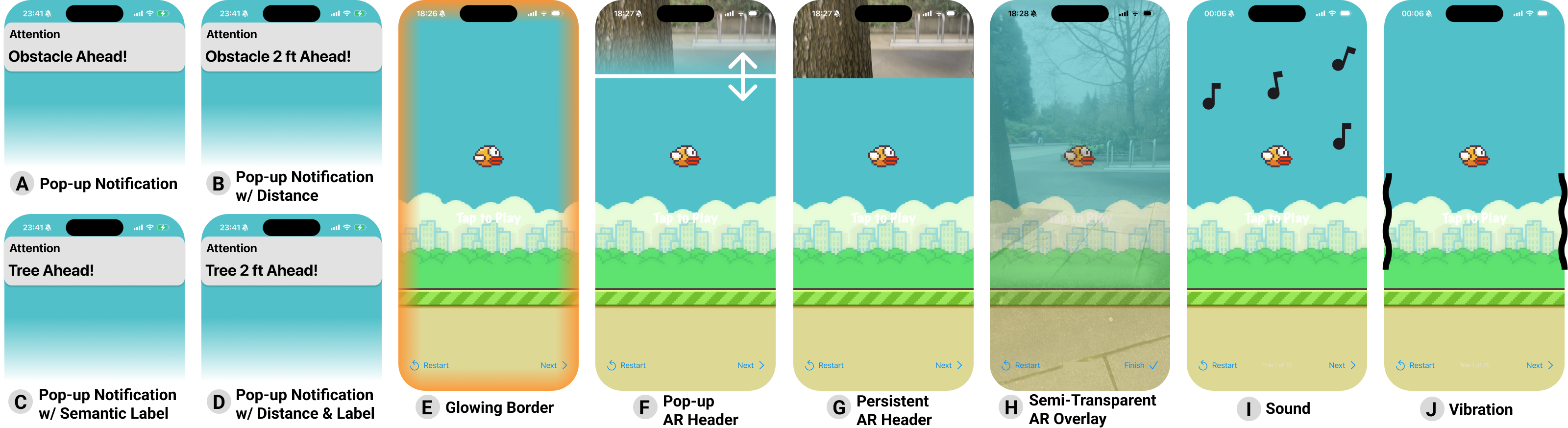}
    \caption{\sysname{} obstacle cue designs.}
    \label{fig:cue_designs}
\end{figure*}

\sysname{} generates various outputs: binary signal indicating obstacle presence, distance to the closest obstacle, semantic label of the obstacle, and forward-facing camera view. Leveraging these data streams, \sysname{} integrates multiple alert designs directly on top of any phone application to warn users of upcoming hazards:
\setlist{nolistsep}
\begin{enumerate}

\item \textit{Pop-up Notifications} (\autoref{fig:cue_designs}A--D).
A standard alert conveying binary, distance, and/or semantic information, auto-dismissing after one second of no further detection.

\item \textit{Glowing Border} (\autoref{fig:cue_designs}E). A colored light rim around the screen perimeter to subtly signal obstacle presence.

\item \textit{AR Header} (\autoref{fig:cue_designs}F--G). A live, forward-facing camera feed at the top of the screen, either as a pop-up triggered by detection or a persistent display.

\item \textit{Semi-Transparent AR Overlay} (\autoref{fig:cue_designs}H). Overlays the forward-facing camera view at 65\% opacity directly onto the primary content, creating a see-through effect.

\item \textit{Non-Visual Cues} (\autoref{fig:cue_designs}I--J). Auditory and haptic feedback that scales in intensity as distance to obstacle decreases.
\end{enumerate}

\section{User Study}


We conducted a within-subjects study evaluating \sysname{} along two dimensions: (1) the objective and perceived accuracy of the sensing pipelines, and (2) the helpfulness of various obstacle cue designs. Each two-hour session included a pre-study questionnaire, tutorial, 16 obstacle course trials across indoor and outdoor environments, and a post-study interview.

\subsection{Participants}

We recruited 11 adult participants (P1–P11; 4 male, 7 female) with an average age of 27.1 ($SD = 6.6$).
All reported using their phone at least once a day while walking.
We excluded individuals with visual, auditory, or mobility impairments.

\subsection{Design Apparatus}

\begin{figure}[h]
    \centering
    \includegraphics[width=\columnwidth]{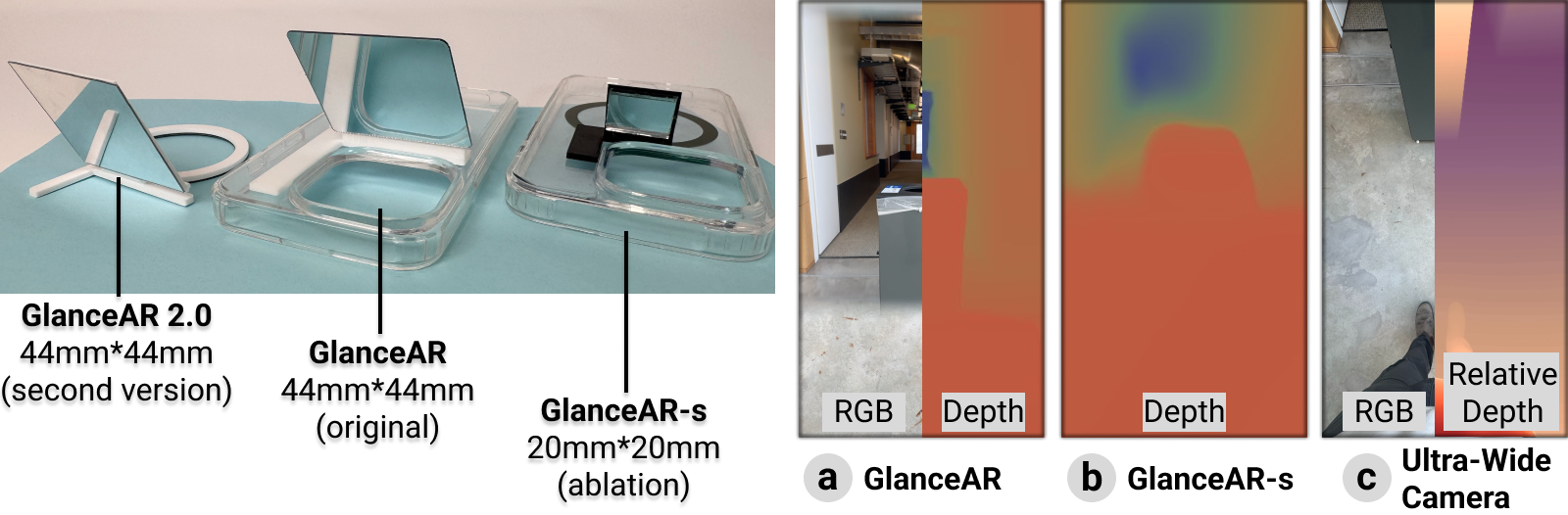}
    \caption{\textit{Left:} \sysname{} 2.0 reflector (see \autoref{sec:second}), \sysname{} reflector, and \sysname{}-s reflector.
    \textit{Right:} Side-by-side comparison of the sensor data of the same scene across \sysname{}, \sysname{}-s, and attachment-free baseline.}
    \label{fig:relative_depth}
    \vspace{-0.2cm}
\end{figure}

We compare \sysname{} against two additional sensing pipelines:

\textbf{LiDAR Ablation.}
To evaluate a privacy-preserving alternative that avoids capturing forward-facing RGB imagery, we test a reduced configuration of \sysname{}, \textit{\sysname{}-s}, that redirects only the LiDAR sensor using a smaller mirror attachment (\autoref{fig:relative_depth}, left 3).
This configuration only covers the LiDAR's narrow FoV ($60^{\circ} \times 48^{\circ}$) and relies on thresholding the metric depth to the closest detected point (\autoref{fig:relative_depth}, right b).

\textbf{Technical Baseline.}
To establish an attachment-free baseline, we evaluate obstacle detection using the iPhone's built-in ultra-wide camera (0.5x, $120^{\circ}$ FoV). Even in a natural downward-facing orientation, this wide field of view captures forward obstacles in the upper portion of the frame (\autoref{fig:relative_depth}, right c).
We define our Region of Interest (ROI) as the top 50\% of the frame in height and width, corresponding to approximately 1.5 m ahead.

To estimate depth, we run \textit{Depth Anything V2}~\cite{yangDepthAnythingV22024} small via Core ML on the uncorrected video feed, achieving 18 FPS on an iPhone 15 Pro. Finally, to differentiate vertical obstacles jutting upward from the horizontal ground plane, we compute the vertical depth gradient within the ROI using finite differences and apply adaptive thresholding to identify strong depth discontinuities.

\subsection{Procedures}
Participants played \textit{Flappy Bird}~\cite{flappybird} as a continuous visual distraction task while the detection pipeline ran simultaneously on an experiment phone. For \sysname{} and \sysname{}-s, the reflector was angled at 60$^{\circ}$, with alert and VLM-trigger thresholds of 1 m and 1.5 m respectively (parameters determined through pilot studies).
Participants first completed a brief tutorial to establish a comfortable posture for holding the phone with the attachment.

\textbf{Indoor trials.}
The indoor trials aimed to assess \sysname{} across three different sensing pipelines: (1) \sysname{}, (2) \sysname{}-s (ablation), and (3) ultra-wide camera-only (baseline).

The indoor course contained 17 obstacles across 10 categories: pedestrians, chairs, tables, trash bins, backpacks, boxes, doors, railings, walls, and elevators. Participants completed six trials (two per sensing pipeline), each using pop-up notifications---the most familiar alert format---displaying as much information as each condition could produce (e.g., \sysname{}-s cannot show obstacle name).

Participants navigated a round trip between two points while aiming to maximize their game score.
We recorded the number of \sysname{} alerts, and used a front-facing camera to track how often participants looked up from the screen.

\textbf{Outdoor trials.}
The outdoor trials evaluated \sysname{} in more naturalistic conditions across the full suite of obstacle cue designs.
We conducted these trials on a preselected campus sidewalk, featuring 20 obstacles spanning 7 categories: pedestrians, poles, trees, traffic signs, mailboxes, bicycles, and parked cars.

Participants again navigated between two points across 10 cue designs (one trial each, as described in \autoref{section:all_cues} and \autoref{fig:cue_designs}). All outdoor trials used the \sysname{} reflector. As before, we recorded alert counts and user look-ups.

\section{Results}
We report results across three dimensions: system sensing performance, participant preferences for obstacle cue designs, and qualitative findings on comfort, trust, privacy, and safety.

\subsection{Measuring System Performance}
To establish baseline capabilities, we first conducted controlled technical evaluations of all three sensing methods using the identical system configuration and indoor course as the user study, but maintained a constant phone tilt of $70^{\circ}$ and height of 1.2 m.
We report both objective accuracy (averaged over three trials per method) alongside the participants' self-reported perceived accuracy (averaged across all 11 participants), in \autoref{tab:system_performance}A.

\sysname{} and \sysname{}-s delivered strong, on-par performance, with the only failure cases being low-profile obstacles such as low tables and ground-level bags. Participants echoed this, rating both systems at 4.0/5, with 9 out of 11 noting low-lying objects as the reason for non-perfect scores. The attachment-free baseline showed lower reliability---higher false positives and reduced precision/recall---with 6 out of 11 participants reporting difficulty detecting large, uniform-textured surfaces like elevator doors.

Despite these accuracy differences, all three methods generated comparable alert volumes and participant look-up frequencies (\autoref{tab:system_performance}B), suggesting that sensing modality alone did not alter glancing-up behavior.

\begin{table}[ht]
\centering
\caption{Summary of system performance and behavioral metrics (mean and SD) for 17 indoor obstacles. Perceived accuracy was rated on a scale of 1 to 5.
}
\label{tab:system_performance}

\footnotesize\textbf{A: Objective and Participant Perceived Accuracy} \\
\footnotesize
\begin{tabular}{@{}l|cccc c@{}}
\toprule
\textbf{Method} &\textbf{TP} & \textbf{FP} & \textbf{Precision} & \textbf{Recall}& \textbf{Perceived Acc.} \\
\midrule
\textit{GlanceAR} & 15.3 (0.6) & 0 (0) & 1.0 (0) & 0.92 (0.03) & 4.0 (1.0) \\
\textit{GlanceAR-s} & 16 (0) & 0 (0) & 1.0 (0) & 0.94 (0) & 4.0 (1.1) \\
Baseline & 11.3 (1.5) & 8.7 (3.2) & 0.65 (0.13) & 0.72 (0.04) & 3.3 (0.8) \\
\bottomrule
\end{tabular}
\vspace{0.1cm}

\textbf{B: Alert and Participant Look-up Frequencies during Trials} \\
\begin{tabular}{@{}l|ccc c@{}}
\toprule
\textbf{Method} & \textbf{Alerts} & \textbf{Look-ups} \\
\midrule
\textit{GlanceAR} & 14.8 (3.3) & 7.2 (6.4)\\
\textit{GlanceAR-s} & 15.1 (3.5) & 7.0 (7.2)\\
Baseline & 15.0 (4.2) & 8.0 (9.0)\\
\bottomrule
\end{tabular}
\vspace{-0.2cm}
\end{table}

\subsection{Obstacle Cue Designs}

Participants ranked their top three preferred obstacle cue designs.
Overall, users strongly favored low-information designs that signal obstacle presence without conveying distance or obstacle type.
The cues most frequently included in participants' top three were vibration (7/11), sound (6/11), plain pop-up notifications (5/11), and glowing border (4/11). In contrast, richer cues such as notifications with semantic label (3/11), persistent AR header (3/11), notifications with distance (2/11), and semi-transparent AR overlay (1/11) were chosen less often.
As P4 noted, "\textit{I can't pay attention to read details; I just need to know something is there.}" Distance values were particularly unpopular among the majority of users, as participants found the dynamic numbers difficult to parse while distracted by the game (P3, P4, P8).



There were some exceptions to these broader trends. P7 ranked pop-up notifications with distance first, explaining that seeing the remaining distance to an obstacle reduced anxiety by signaling that there was still time to prepare and respond. Another exception was P11, who ranked semantic pop-up notifications with semantic label first. They noted that the labels reminded them of \textit{"road signs like `Deer Ahead'"} and valued the direct efficiency of a textual summary.

\subsection{Comfort, Trust, Privacy and Safety}
Beyond performance and cue preferences, we assessed participants' subjective experience with \sysname{} across four dimensions.

\textbf{Comfort.}
8 out of 11 participants reported neutral to favorable comfort levels with the \sysname{} attachment, while all participants reported positive comfort levels for \sysname{}-s and the attachment-free baseline.
2 participants noted the attachment acted as an ergonomic grip where their fingers naturally rested.

\textbf{Trust.} Participants placed limited trust in the alerts across all three pipelines, particularly after experiencing inconsistent alerts and false negatives. 
\textit{"[I] would still look up frequently when walking"} (P6); \textit{"Anything could go wrong"} (P3).
Trust improved when participants could visually verify detections via the live camera feed (P5), or when semantic labels signaled genuine scene understanding rather than arbitrary alerts (P11).

\textbf{Privacy.} 9 out of 11 participants agreed they would feel uncomfortable seeing someone use \sysname{} publicly, and 4 preferred \sysname{}-s over \sysname{} because of the absence of a forward-facing camera.
Notably, participants expressed minimal concern about their \textit{own} data, particularly when processing remained on-device; instead, privacy concerns centered on \textit{bystander} perception and potential misuse.
P9 raised a specific concern that reflective attachments could be co-opted by sexual harassers to discreetly photograph others.

\textbf{Safety.} 5 participants felt safer with \sysname{}, 3 reported no change, and the remaining 3 felt less safe, primarily stating that the system introduced additional distraction.
Several reported strong pre-existing peripheral awareness---P3 asserted they \textit{"could navigate using peripheral vision without looking up."}
In contrast, P11, who felt safer, explained that it \textit{"feels like I have another thing that interprets the environment, not alone by myself".}
P8 noted that \sysname{} could be especially valuable in high-mobility contexts such as running or biking, where peripheral monitoring is harder to maintain.







\section{\sysname{} 2.0}
\label{sec:second}

Based on participant feedback, we developed \sysname{} 2.0 to improve the everyday usability of the initial prototype (\autoref{fig:relative_depth}, left 1). 
Inspired by participants' suggestions to mimic collapsible phone grips (e.g., PopSockets), the second version utilizes a MagSafe-compatible ring for quick attachment and detachment while maintaining optical alignment.
Furthermore, \sysname{} 2.0 incorporates a hinge mechanism. This design not only allows the reflector to collapse completely flat for easy pocketability, but also enables users to dynamically fine-tune the mirror's angle, making it possible to optimize the forward-facing field of view across varying user heights and phone-holding postures.
\section{Discussion}
Surprisingly, our findings indicate that accuracy mattered less than cue design in shaping participant behavior, with look-up rates comparable across all three sensing pipelines despite large differences in objective accuracy. We unpack these dynamics below.

\textbf{Risk of Automation Complacency.}
Although \sysname{} is an early prototype, there is a concern that higher-accuracy future versions could cause over-reliance and introduce new safety risks.
\textit{"If you have it, people tend to rely on it. As long as it makes mistakes, it can cause huge bad things for people, like being hit by a car"} (P4).
P5 described a shift toward \textit{"letting go of the need to control the situation"} when using pop-up/persistent AR overlay cues.
These observations suggest that as reliability improves, always-on mobile safety systems risk encouraging an "autopilot" behavior in which users delegate hazard avoidance to the AI. Cue designs that prompt phone users to \textit{look up and verify}, rather than rendering hazard information on-screen, may counteract this tendency.

\textbf{Attention as a Scarce Resource.}
We observed a tension between prompting users to look up for safety and preserving immersion in the primary task.
Some participants seamlessly context-switched between a quick glance and the on-screen task; others found that breaking visual contact derailed in-game performance and preferred persistent or information-rich cues.
Yet both groups tolerated imperfect accuracy and false positives.
We attribute this to the low cost of false alarms.
Because users can verify a hazard with a quick glance, the system does not need flawless accuracy to be useful---it should act as a low-friction trigger to re-engage natural vigilance.

\textbf{Bystander Privacy and Social Acceptability.} 
Capturing the forward environment from a natural downward posture raises privacy concerns primarily for bystanders.
A raised phone traditionally signals recording intent; \sysname{}'s passive reflector bypasses this social norm.
Future work must explore how low-cost physical augmentations reshape the social contract of mobile sensing.

\textbf{Limitations and Future Work.} \sysname{} remains constrained in several ways.
Building on the manual hinge introduced in \sysname{} 2.0, future designs should also explore motorized actuation to automatically adjust the reflector's angle, accommodating varying user heights and dynamic shifts in posture.
Additionally, future iterations should replace our bulky external planar mirror with refractive or periscope optics to minimize the form factor and increase robustness in different lighting conditions.
The concept of an "obstacle" is also context-dependent; a static object may be critical on a fast sidewalk but irrelevant in a slow crowd. Participants explicitly requested dynamic alert thresholds tuned to context (e.g., approaching an intersection vs. strolling in a park), and the scene-level understanding capabilities of VLMs offer a promising path toward this adaptation, which we begin to explore in this work by enabling an open vocabulary of obstacle types.

Despite these constraints, \sysname{} demonstrates, through both user study and technical evaluation, that effective obstacle awareness can operate alongside how people naturally hold their devices and interact with pre-existing apps.
We hope this initial exploration inspires always-on mobile safety systems where low-cost, passive sensor augmentations, intuitive cues, and on-device AI work in tandem to support users across diverse, safety-critical contexts.

\bibliographystyle{ACM-Reference-Format}
\bibliography{references}



\end{document}